\documentclass[sigconf,dvipsnames]{acmart}
\AtBeginDocument{%
  }

\usepackage[final]{listings}
\usepackage{caption}
\usepackage{algorithm}
\usepackage[noend]{algpseudocode}
\usepackage{cleveref}

\lstdefinelanguage{JavaScript}{
  keywords={break, case, catch, const, continue, debugger, default, delete, do, else, export, for, function, if, import, in, instanceof, let, new, return, super, switch, this, throw, try, typeof, var, void, while, with, yield},
  keywordstyle=\color{blue}\bfseries,
  ndkeywords={class, export, boolean, throw, implements, import, this},
  ndkeywordstyle=\color{magenta}\bfseries,
  identifierstyle=\color{black},
  sensitive=false,
  comment=[l]//,
  morecomment=[s]{/*}{*/},
  commentstyle=\color{gray}\ttfamily,
  stringstyle=\color{Rhodamine}\ttfamily,
  morestring=[b]',
  morestring=[b]",
  literate=
  {"term"}{{\textcolor{Green}{"term"}}}{6}
  {"source"}{{\textcolor{Green}{"source"}}}{8}
  {"applicable_to"}{{\textcolor{Green}{"applicable\_to"}}}{15}
  {"verification"}{{\textcolor{Green}{"verification"}}}{14}
  {"justification"}{{\textcolor{Green}{"justification"}}}{15}
  {"possible_accountability_checks"}{{\textcolor{Green}{"possible\_accountability\_checks"}}}{30}
}

\lstdefinestyle{mystyle}{
  backgroundcolor=\color{white},
  basicstyle=\ttfamily\footnotesize,
  breaklines=true,
  captionpos=b,
  numbers=left,
  numberstyle=\tiny\color{gray},
  numbersep=5pt,
  showstringspaces=false,
  tabsize=2,
}

\lstdefinestyle{txtfile}{
  basicstyle=\ttfamily\footnotesize,
  breaklines=true,
  captionpos=b,
}

\def\framework{\texttt{Terminators}}

\copyrightyear{2025}
\acmYear{2025}
\setcopyright{acmlicensed}\acmConference[Websci Companion '25]{Proceedings of the 17th ACM Web Science Conference Companion 2025}{May 20--24, 2025}{New Brunswick, NJ, USA}
\acmBooktitle{Proceedings of the 17th ACM Web Science Conference Companion 2025 (Websci Companion '25), May 20--24, 2025, New Brunswick, NJ, USA}
\acmDOI{10.1145/3720554.3736183}
\acmISBN{979-8-4007-1535-8/2025/05}

\begin{document}

\title{Terminators: Terms of Service Parsing and Auditing Agents}

\author{Maruf Ahmed Mridul}
\authornote{Both authors contributed equally to this research.}
\email{mridum@rpi.edu}
\affiliation{%
	\institution{Rensselaer Polytechnic Institute}
	\city{Troy}
	\state{New York}
	\country{USA}
}

\author{Inwon Kang}
\authornotemark[1]
\email{kangi@rpi.edu}
\affiliation{%
	\institution{Rensselaer Polytechnic Institute}
	\city{Troy}
	\state{New York}
	\country{USA}
}

\author{Oshani Seneviratne}
\orcid{0000-0001-8518-917X}
\email{senevo@rpi.edu}
\affiliation{%
	\institution{Rensselaer Polytechnic Institute}
	\city{Troy}
	\state{New York}
	\country{USA}
}

% \affiliation{%
% 	\institution{Inria Paris-Rocquencourt}
% 	\city{Rocquencourt}
% 	\country{France}
% }


\begin{abstract}
	Terms of Service (ToS) documents are often lengthy and written in complex
	legal language, making them difficult for users to read and understand. To address this challenge, we propose \textbf{Terminators}, a modular agentic framework that leverages large language models (LLMs) to parse and audit ToS documents. Rather than treating ToS understanding as a black-box summarization problem, \textbf{Terminators} breaks the task down to three interpretable steps: term extraction, verification, and accountability planning. We demonstrate the effectiveness of our method on the OpenAI ToS using GPT-4o, highlighting strategies to minimize hallucinations and maximize auditability.
	% Our results suggest that structured, agent-based LLM workflows
	% can enhance both the usability and enforceability of complex legal documents,
	% with potential applications in user-facing compliance tools and automated
	% policy audits.
	Our results suggest that structured, agent-based LLM workflows can enhance both the usability and enforceability of complex legal documents. By translating opaque terms into actionable, verifiable components, Terminators promotes ethical use of web content by enabling greater transparency, empowering users to understand their digital rights, and supporting automated policy audits for regulatory or civic oversight.
\end{abstract}



\maketitle

\section{Introduction}

When entering a website, users are often presented with a long and complex document describing the Terms of Service (ToS). These documents are often written in a legal language, making them difficult to understand for the average user. Such complexity often leads to the users not even reading the ToS at all, either because they are too long or because they are written in a language that is difficult to understand~\cite{moallem2018you}. This can lead to situations where users unknowingly agree to terms that they do not fully understand, which can have serious consequences as the user is often bound by the terms and conditions outlined in the document. Thus, it is crucial that the users can understand the meaning and nuances of this document. In recent years, there have been efforts to require such web services to be more transparent and explicit about their usage of the user's data, as well as giving more control to the users~\cite{gdpr,ccpa,cpa}. General Data Protection Regulation (GDPR), for example, requires that users are informed about the data that is being collected, how it is being used, and who it is being shared with. In addition, it also requires that apps and service providers are audited to ensure that they are in compliance with the regulations~\cite{gdpr}. However, as the complexity of modern software systems grow, so can the difficulties of conducting such audits.
% removed ,ucpa from the references. 

Recent advances in large language models (LLMs) offer promising opportunities to address these challenges. In particular, these models have shown great strength in handling unstructured documents, such as the ability to parse verbose documents to automate summarization and analysis of the content~\cite{van2024adapted,radford2018improving,liu2023learning,zhang2024benchmarking}. These capabilities can help the everyday user to better understand complex documents by providing a more layman-friendly version of document. In fact, this functionality can be further extended to automate many aspects in handling complex documents like ToS documents, starting from parsing and analyzing the terms and even planning how to verify that the terms are being upheld. However, when designing such a system, we have to be mindful of the drawbacks of LLMs -- namely, their unreliability when the task involves handling \textit{facts} from the input document, a phenomena often referred to as \textit{hallucination}~\cite{huang2025survey,xu2024hallucination,tang2023evaluating}. This introduces a new challenge when considering incorporating language models for handling documents as fact-heavy as ToS documents: How can we ensure the automation of the process is conducted \textit{correctly}?

In this work, we explore the potential -- and limitations -- of using LLMs for interpreting and building on top of digital service agreements. We propose an agentic framework that leverages LLMs to parse ToS documents into a set of auditable \textit{terms}. This framework can be used by auditors or users to verify the truthfulness of the service's ToS documents, or to verify whether the service is being used correctly within the bounds of the ToS.
% Maruf: addressing reivew 5
Reflecting the diverse and often complex nature of online agreements, this framework is designed to be general enough to handle any ToS documents, primarily those of digital platforms and web services.
% We also discuss the implications of using LLMs in this context, including the need for accountability and traceability in the auditing process.

\section{Related Work}

Early approaches to ToS analysis focused on machine learning-based clause classification, such as sentence-level detection of unfair clauses in ToS documents~\cite{lippi2019claudette} or classifying both category and fairness of clauses~\cite{guarino2021machine,kokila2024machine}.
Beyond classification, Karanicolas~\cite{karanicolas2021too} examined the interpretability challenges of boilerplate ToS, highlighting the structural mismatch between user-facing policies and enforceable legal terms. In parallel, Azose~\cite{azose2023generating} introduced the idea of using LLMs to auto-summarize ToS in real-time, improving accessibility for end users.

With the emergence of LLMs, recent work has focused on more comprehensive ToS analysis, such as extracting and identifying key concepts~\cite{soneji2024demystifying} or evaluating LLMs on their question-answering ability~\cite{frasheri2024llm}, to even proposing an end-to-end comliance automation~\cite{hassani2024rethinking}. However, it is worth noting that~\citet{frasheri2024llm} report limited effectiveness of LLMs in the vanilla (unstructured) question-answering paradigm.

It is also worth noting the existence of a community-based ToS summarization platform\footnote{\url{https://tosdr.org/en}}. While this platform provides a more intuitive explanation of the ToS documents of various popular websites, it is limited to user contributions and it does not provide any mechanism to ensure whether the ToS is being upheld by any parties involved.

\section{Proposed Method}

In this work, we propose \framework{}, a framework that leverages LLMs to parse ToS documents into a set of \textit{semantically whole} chunks. Instead of simply asking the LLM to summarize a long ToS document in one step, we break down the process into a sequence of actions to minimize errors (e.g., \textit{hallucinations} by the model), as well as adding auditability to the process by explicitly defining steps that can be examined by a human expert. 
% inwon: it's a known thing about chunking documents, so best we can do is point that others showed this already.
Previous works~\cite{liu2023lost,asai2023self} have shown that naively providing a long document as a context to an LM can lead to inferior performance compared to when only the relevant ``chunks'' are provided. Inspired by this finding, we aim to leverage the structured nature of ToS documents to break the long document into semantically whole chunks, which can then be provided to the LLM for further processing.
Given a long ToS document, the end goal of \framework{} is to parse it into a series of structured \textit{terms}, which can then be used in downstream applications such as auditing the privacy of the service or ensuring that the user adheres by the service provider's terms. In this work, we define \textit{term} as a clause that defines the relationship between the parties involved -- typically, the user and the service provider. \Cref{lst:json-example} shows an example of how a single term may be structured.

\begin{lstlisting}[language=JavaScript, style=mystyle, caption={Example of a parsed ToS snippet}, label={lst:json-example}]
{
  "term": "This website will not share personally identifying information",
  "source": "WebsiteToS.txt:70-72",
  "applicable_to": ["website"]
},
\end{lstlisting}

\subsection{\framework{}}

\begin{figure}[h]
	\centering
	\includegraphics[width=1\linewidth]{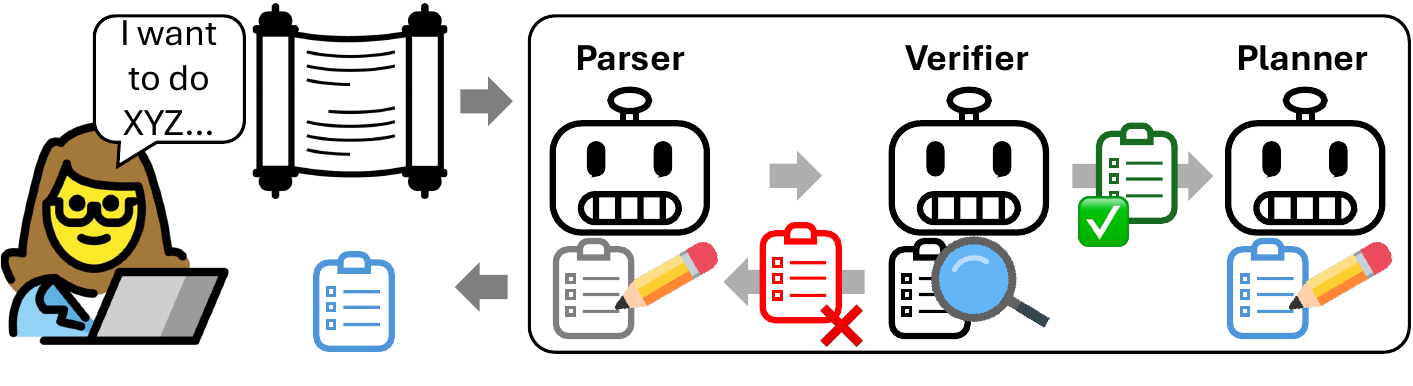}
	% \vspace{-0.3cm}
	\caption{Overview of \framework{}.}
	\label{fig:overview}
	% \vspace{-0.5cm}
\end{figure}

\framework{} is composed of three agents -- term parsing, term verification, and accountability -- that each have their distinct roles. Below we will describe each component in detail. \Cref{fig:overview} shows an overview of how the agents interact in \framework{}.

\subsubsection{Term Parser}

The main task of the term parsing agent is to extract \textit{terms} from the ToS document. A term is defined as a specific, declarative statement that communicates obligations, permissions, restrictions, rights, or responsibilities of either the user or the service provider. To avoid possible hallucinations, we require that the agent supply the \textit{source} of the term when parsing. We also specify that the extracted term must follow the format as seen in~\Cref{lst:json-example}.

\subsubsection{Term Verifier}

Once the terms have been extracted from the input document, they are passed to the \textbf{verifier agent}, which evaluates whether the term is faithful to its source. The main role of the verifier is to check whether each term is indeed: 1) a valid term, 2) is sourced from the ToS document, 3) and is correctly cited from the source. The verifier may respond using the following labels:
\begin{itemize}
	\item \textbf{Supported}: The term is well-supported by the source text, either through direct mention or through clear contextual implication.
	\item \textbf{Contradicted}: The term directly conflicts with the source or misrepresents what it states.
	\item \textbf{Unverifiable}: The term is not explicitly supported or contradicted by the source; the source is vague, incomplete, or silent on the term. Or, the source is not relevant to the term.
\end{itemize}
In addition, the verifier may also provide a free-text justification for why the label was assigned. This is useful for both a human expert who may wish to debug the process, or the next agent that will be triggered to amend the incorrect term. If the label is anything other than \texttt{Supported}, a new term parsing agent is triggered to review the ToS document for a more appropriate source span corresponding to the original term. If a better source is found, we update the term's source; otherwise, we discard the term.

\subsubsection{Scenario-Aware Accountability Planner}

Finally, once the terms have been processed and verified by the parser and verifier agents, we introduce an agent that \textit{plans} on how to verify the accountability of the terms. The objective of this agent is to plan a set of actionable items that can be used to check whether the terms that are described in the original ToS document are being upheld, thus checking for \textit{accountability}. We adopt a scenario-aware approach to improve the relevance and effectiveness of the generated accountability check instructions. Rather than producing generic or abstract checks, we condition the planner on a realistic user scenario that reflects how the system is actually used.
% Maruf: addressing review 6
Additionally, the planner can be extended to account for regional ToS requirements, such as the General Data Protection Regulation (GDPR)~\cite{gdpr} in the EU and the California Consumer Privacy Act (CCPA)~\cite{ccpa}, enabling geographically appropriate accountability checks.
This allows the resulting instructions to be both context-specific and actionable from the user's perspective.

\subsection{Preliminary Experiments}
\label{sec:observations}

To test the applicability of \framework{} in real-life scenarios, we experiment using GPT-4o\footnote{\url{https://platform.openai.com/docs/models/gpt-4o}} through ChatGPT-4o with the publicly available OpenAI ToS document~\cite{openaiToS}.

\subsubsection{Parser Agent}

Despite clear line breaks in the ToS, the parser agent often hallucinates incorrect source line numbers. While the extracted terms are usually valid, source locations can be imprecise. We find that \textit{explicitly adding line numbers} significantly improves source accuracy. In addition, we find that the parser agent occasionally fails to extract all valid terms when processing the full ToS document in a single prompt. To remedy this, we experimented with a parallelized approach—running multiple parser agents concurrently and merging their outputs -- but found that this approach still resulted in missed terms. However, breaking the ToS into smaller sections or paragraphs markedly improves term coverage. As a result, we adopt a section-by-section parsing strategy.

For instance, when the full OpenAI ToS is input, GPT-4o identifies the terms in Listing~\ref*{lst:identified-from-whole-ToS} for lines 106–117 (shown in Listing~\ref*{lst:openai_ToS}). But when only this paragraph is used as input, the agent extracts a richer set of terms (Listing~\ref*{lst:identified-from-a-para}), demonstrating the advantage of targeted, paragraph-level prompts.

\begin{lstlisting}[style=txtfile, caption={Excerpt from OpenAI's ToS}, label={lst:openai_ToS}]
106: When you use our Services you understand and agree:
107:
108: Output may not always be accurate. You should not rely on Output from our Services as a sole
109: source of truth or factual information, or as a substitute for professional advice.
110: You must evaluate Output for accuracy and appropriateness for your use case, including using
111: human review as appropriate, before using or sharing Output from the Services.
112: You must not use any Output relating to a person for any purpose that could have a legal or
113: material impact on that person, such as making credit, educational, employment, housing,
114: insurance, legal, medical, or other important decisions about them.
115: Our Services may provide incomplete, incorrect, or offensive Output that does not represent
116: OpenAI's views. If Output references any third party products or services, it doesn't mean the
117: third party endorses or is affiliated with OpenAI.
\end{lstlisting}

\begin{lstlisting}[language=JavaScript, style=mystyle, caption={Example of terms identified using the full ToS document}, label={lst:identified-from-whole-ToS}]
{
  "term": "Users must not rely on Output as a sole source of truth or substitute for professional advice.",
  "source": "OpenAI_ToS.txt:108-109",
  "applicable_to": ["user"]
},
{
  "term": "Users must not use any Output for decisions with legal or material impact on individuals, such as employment or medical decisions.",
  "source": "OpenAI_ToS.txt:112-114",
  "applicable_to": ["user"]
}
\end{lstlisting}

\begin{lstlisting}[language=JavaScript, style=mystyle, caption={Example of terms identified using a single paragraph}, label={lst:identified-from-a-para}]
{
  "term": "Users must not rely on Output from the Services as a sole source of truth, factual information, or a substitute for professional advice.",
  "source": "OpenAI_ToS.txt:108-109",
  "applicable_to": ["user"]
},
{
  "term": "Users must evaluate Output for accuracy and appropriateness for their use case, including using human review as appropriate, before using or sharing Output.",
  "source": "OpenAI_ToS.txt:110-111",
  "applicable_to": ["user"]
},
{
  "term": "Users must not use any Output relating to a person for any purpose that could have a legal or material impact on that person, including making credit, educational, employment, housing, insurance, legal, or medical decisions.",
  "source": "OpenAI_ToS.txt:112-114",
  "applicable_to": ["user"]
},
{
  "term": "The Services may produce Output that is incomplete, incorrect, or offensive and does not represent OpenAI's views.",
  "source": "OpenAI_ToS.txt:115",
  "applicable_to": ["OpenAI"]
}
\end{lstlisting}

Since ToS sentences are often declarative, the agent tends to extract nearly all as terms, resulting in broad outputs. To focus extraction, one can specify target aspects (e.g., data sharing, privacy). For example, when asked to extract terms related to ``user rights and responsibilities,'' the last term in Listing~\ref*{lst:identified-from-a-para} is omitted, as it does not match the aspect desired in that context. In contrast, a vague aspect like ``output usage and consequences'' causes the agent to include even non-user-facing disclaimers (e.g., line 115). This suggests that poorly scoped aspects can lead to noisy results due to overgeneralization.

\subsubsection{Terms Verifier}

The verifier agent reliably assesses whether a term is supported by its cited source. Typically, the parser agent also identifies the correct source span—especially when the ToS document includes explicit line numbers, which significantly enhances source extraction accuracy.

Without line numbers, occasional mismatches occur between a term and its cited source. In such cases, the verifier correctly flags inconsistencies. For instance, consider this parser-extracted term:

\begin{lstlisting}[language=JavaScript, style=mystyle]
{
  "term": "You may not attempt to reverse engineer, decompile or discover the source code or underlying components of the Services.",
  "source": "OpenAI_ToS_Raw.txt:28",
  "applicable_to": ["user"]
}
\end{lstlisting}

The actual source text at line 28 is: \textit{"Modify, copy, lease, sell or distribute any of our Services."}

When passed to the verifier agent, the system returned:

\begin{lstlisting}[language=JavaScript, style=mystyle]
{
  "verification": "Unverifiable",
  "justification": "The source text prohibits modifying, copying, leasing, selling, or distributing the Services, but it does not mention reverse engineering, decompiling, or discovering source code. Therefore, the term cannot be verified based solely on the provided source passage."
}
\end{lstlisting}

This example shows the verifier performs as expected—detecting unsupported terms—and highlights the parser's need for precise source selection. When ToS documents include line numbers, all sampled terms in our evaluation were correctly linked and verified as \texttt{Supported}.

\subsubsection{Accountability Planner}

Generating multiple accountability checks provides broader coverage, especially when the service's inner workings are opaque (e.g., closed-source proprietary websites). In such cases, a single verification method may be insufficient, so a tailored set of checks increases the chances of capturing relevant and verifiable mechanisms.

\textbf{Scenario.} \textit{``I am a university student using this AI service for coursework and research—asking it to explain concepts, summarize articles, or answer factual questions. I want to ensure the service clearly communicates the limitations of its output to avoid using inaccurate information in assignments or reports.''}

Consider the term: \textit{``Users must not rely on Output from the Services as a sole source of truth, factual information, or a substitute for professional advice.''}

The accountability planner generates the following checks (Listing~\ref{lst:accountability-student}) that a student could reasonably observe, test, or evaluate:

\begin{lstlisting}[language=JavaScript, style=mystyle, caption={Accountability Checks from a Student's Perspective}, label={lst:accountability-student}]
{
  "possible_accountability_checks": [
    "Check if the service displays clear disclaimers before or after generating answers about academic, scientific, or factual content.",
    "Attempt to use the AI to summarize or explain a scholarly article and observe whether the service warns about the need for external verification.",
    "Search the website's documentation or FAQ to find guidance for students or researchers about verifying factual Output.",
    "Evaluate whether the service includes citations, confidence scores, or 'double-check' prompts when presenting factual terms.",
    "Test if queries asking for legal, medical, or academic conclusions are flagged or responded to with cautionary language."
  ]
}
\end{lstlisting}

This scenario-aware approach aligns accountability checks with the user's context, promoting observable safeguards and enhancing the practical auditability of AI services.

\section{Discussion \& Further Application}
\label{sec:discussion}

In this work, we propose \framework{}, a framework of LLM agents that can be used to parse a verbose ToS document into a set of concise and well-formatted mini-documents. We showcase an example of \framework{} by implementing with GPT-4o and parsing OpenAI's ToS document, and discuss our findings on strategies that seem to work the best for each agent.

While we focus on the \textit{verification} of ToS document in our work, it is important to note that its application is not limited to such cases. In fact, one of the  most natural use case for \framework{} is to use it as a part of a query engine to \textit{interact} with a ToS document.
% Maruf: addressing review 7
This interaction can include targeted checks for specific terms, such as whether cookies are transferred to third parties or whether personal data is shared for advertising purposes.
This is enabled by the fact that \framework{} organizes the document at the \textit{term} level, accompanied with the appropriate citation in the original document. Compared to the commonly used retrieval-augmented generation (RAG) approaches that rely on breaking the text into fixed-length blocks, our approach can greatly enhance the quality of context provided to the language model when faced with a user query.
For example, a researcher working on a project that uses data from a service may want to ensure that their work adheres to the service's ToS. In this case, the researcher can use \framework{} to query the ToS document for specific terms related to data usage and sharing. The system can then provide the relevant terms along with their sources, allowing the researcher to quickly assess whether their work is in compliance with the ToS.

In this work, we aim to provide a general-use framework to deal with website ToS documents that can be applied in different scenarios, with a focus on parsing the document into a more computable format.
% Maruf: addressing review 4
% \textcolor{red}{However, while newer reasoning LLMs such as OpenAI's GPT-o3 have improved in accuracy and reduced hallucinations for such tasks, this framework provides additional benefits. It produces structured, context-rich outputs directly usable for downstream tasks like compliance checks, with each term explicitly linked to its source for better traceability. Its modular design also supports flexible, scenario-aware workflows, allowing customization based on specific application needs, which can be challenging for generic LLMs to handle without extensive fine-tuning.}
% inwon: we are not proposing a new LM, so we shouldn't try to compare i think. I think instead we can say our framework will *benefit further* from these nice LMs.
With the release of newer models such as OpenAI's GPT-o3 that are equipped with even better reasoning capabilities and context size, we expect that \framework{} will become even more effective. The main spirit of \framework{} lies in the organization of the document -- and by using a more capable LLM, we can expect to construct even better accountability checks or more complex queries using \framework{}

% Future works may expand this work by including more advanced features, such as the ability to parse a more sophisticated relationship between entities (e.g., knowledge graphs).
% Maruf: updated the above sentence with the following to address review 1
Future work may expand this framework by incorporating more advanced classification mechanisms, such as organizing verified terms into machine-interpretable structures like knowledge graphs, RDF triples, or JSON-LD. This could involve grouping terms based on their semantic roles (e.g., Obligations, Permissions, User Rights, Data Handling) and defining hierarchical relationships (e.g., "is governed by," "requires consent," "is restricted by") to capture their broader legal context.
% inwon: addressing the empirical evidence review
In this work, we introduce \framework{} as a proof-of-concept framework along with small-scale experiments to demonstrate its efficacy. We aim to provide a foundation for future work to extend upon this work to conduct a more comprehensive design and evaluation.

\bibliographystyle{ACM-Reference-Format}

\footnotesize{\bibliography{references}}

\end{document}